\documentclass[lettersize,journal]{IEEEtran}
\usepackage{amsmath,amsfonts}
\usepackage{algorithmic}
\usepackage{algorithm}
\usepackage{array}
\usepackage[caption=false,font=normalsize,labelfont=sf,textfont=sf]{subfig}
\usepackage{textcomp}
\usepackage{stfloats}
\usepackage{url}
\usepackage{verbatim}
\usepackage{graphicx}
\usepackage{cite}
\usepackage{siunitx}
\usepackage[capitalize,nameinlink]{cleveref}
\crefname{equation}{}{}
\Crefname{equation}{Equation}{Equations}

\usepackage{adjustbox}
\usepackage{nicematrix}
\usepackage{booktabs}
\usepackage{multirow}
\usepackage{threeparttable}

\usepackage{tikz,pgfplots}
\usepgfplotslibrary{groupplots,units}
\usetikzlibrary{patterns}
\pgfplotsset{compat=newest}
\usetikzlibrary{shapes.geometric, arrows, calc, fit, shapes, backgrounds}
\usepgfplotslibrary{colorbrewer}

\newcommand{\BE}{\begin{equation*}\begin{aligned}}
\newcommand{\EE}{\end{aligned}\end{equation*}}

\newcommand{\score}{\mathcal{A}}
\newcommand{\pooldist}{\mathcal{D}}
\newcommand{\cX}{\mathcal{X}}

\usepackage{acronym}
\acrodef{asd}[ASD]{anomalous sound detection}
\acrodef{dtw}[DTW]{dynamic time warping}
\acrodef{eat}[EAT]{efficient audio transformer}
\acrodef{gem}[GeM]{generalized mean}
\acrodef{gwrp}[GWRP]{global weighted ranking pooling}
\acrodef{rdp}[RDP]{relative deviation pooling}
\acrodef{ssl}[SSL]{self-supervised learning}
\acrodef{stft}[STFT]{short-time Fourier transform}

\makeatletter
\newcommand*{\org@overidelabel}{}
\let\org@overridelabel\AC@verridelabel
\renewcommand*{\AC@verridelabel}[1]{%
  \@bsphack
  \protected@write\@auxout{}{\string\AC@undonewlabel{#1@cref}}%
  \org@overridelabel{#1}%
  \@esphack
}%
\makeatother

\usepackage{color}
\definecolor{light}{rgb}{0.5, 0.5, 0.5}

\definecolor{answerblue}{rgb}{0.21,0.37,0.57}

\begin{document}

\title{Temporal Pooling Strategies for Training-Free Anomalous Sound Detection with Self-Supervised Audio Embeddings}

\author{Kevin Wilkinghoff,~\IEEEmembership{Senior Member,~IEEE,} Sarthak Yadav, and Zheng-Hua Tan,~\IEEEmembership{Senior Member,~IEEE}
\thanks{The authors are with the Department of Electronic Systems, Aalborg University, Aalborg, Denmark and Pioneer Centre for Artificial Intelligence, Denmark (e-
mail: kevin.wilkinghoff@ieee.org, sarthaky@es.aau.dk, zt@es.aau.dk.}
}



\maketitle

\begin{abstract}
Training-free \ac{asd} based on pre-trained audio embedding models has recently garnered significant attention, as it enables the detection of anomalous sounds using only normal reference data without task-specific model training or fine-tuning.
However, existing embedding-based approaches almost exclusively rely on temporal mean pooling, leaving temporal pooling in training-free \ac{asd} largely unexplored.
In this paper, we present the first systematic evaluation of temporal pooling strategies for training-free \ac{asd} with pre-trained audio embeddings. We propose \ac{rdp}, an adaptive pooling method that assigns larger weights to embeddings with stronger temporal deviations, investigate feature-wise non-linear aggregation using \ac{gem} pooling, and examine a hybrid combination of both strategies.
Experiments on five benchmark datasets demonstrate that the proposed pooling strategies consistently outperform mean pooling and achieve state-of-the-art performance for training-free \ac{asd}, including results that surpass previously reported trained systems and ensembles on the DCASE2025 \ac{asd} dataset.
\end{abstract}


\acresetall

\section{Introduction}
\IEEEPARstart{S}{emi}-supervised \ac{asd} is the task of distinguishing between normal and anomalous recordings of an acoustic phenomenon, while only having access to normal reference samples representing the meaning of the term \emph{normal} for a particular application.
State-of-the-art \ac{asd} systems typically rely on projecting acoustic signals into a latent embedding space and computing distances to the reference samples, which serve as anomaly scores.

\par
In challenging acoustic conditions, off-the-shelf pre-trained audio embedding models often underperform discriminative methods that are trained using metadata or labeled auxiliary tasks \cite{wilkinghoff2023using,fujimura2025asdkit}.
The reason for this is that discriminative systems can implicitly suppress irrelevant signal components, such as background noise, by focusing on features that are predictive of the target labels, thereby improving robustness in noisy environments \cite{wilkinghoff2024why}.
In contrast, training-free approaches must rely solely on the structure of the acoustic representations themselves, and thus are equally sensitive to both irrelevant signal components as well as possibly subtle target signal components indicating anomalies.

\par
Recently, however, training-free ASD methods based on large-scale pre-trained audio embedding models, i.e., approaches without task-specific model training or fine-tuning, have attracted growing interest \cite{mezza2023zero-shot,saengthong2024deep,wu2025towards,zhang2025echo,fan2025fisher}.
Such approaches offer several advantages: They reduce reliance on domain-specific metadata, generalize better under domain shifts \cite{wilkinghoff2025handling}, avoid the risk of emphasizing metadata-discriminative at the expense of anomaly-discriminative signal components, and can be readily applied for pseudo-labeling or bootstrapping discriminative systems when metadata is scarce \cite{fujimura2025improvements,fujimura2025discriminative,fujimura2026pseudo}.
These properties make training-free \ac{asd} particularly attractive for rapidly deployable and scalable monitoring systems.

\par
Most pre-trained audio embedding models produce sequences of frame-level embeddings whose length depends on the input duration.
Direct comparison of such variable-length sequences is computationally expensive and memory-intensive, as it requires storing and matching full embedding trajectories.
To enable efficient similarity computation, temporal pooling is therefore used to aggregate sequences into fixed-dimensional representations that can be compared using simple distance measures such as Euclidean or cosine distance.

\par
Despite the central role of temporal pooling in embedding-based training-free \ac{asd}, its impact has remained largely unexplored. Existing approaches almost exclusively rely on temporal mean pooling, although temporal pooling is one of the few architectural components that can be modified without introducing supervision.
This is particularly relevant for anomaly detection, where informative events are often brief and localized rather than uniformly distributed over time.
Mean pooling implicitly assumes that all temporal segments are equally informative and therefore contribute equally to the pooled representation. While this may reduce the influence of background noise and other sources of variability, it may also diminish the contribution of brief anomalous events when they are averaged with predominantly normal segments, making anomalous recordings more difficult to distinguish from normal ones.

\IEEEpubidadjcol

\par
This motivates temporal pooling strategies that explicitly account for the distribution of temporal deviations within an embedding sequence. Based on this insight, we propose \ac{rdp}, a novel training-free pooling strategy that assigns larger weights to frames exhibiting stronger temporal deviations from the sequence average. We further investigate feature-wise non-linear aggregation using \ac{gem} pooling \cite{radenovic2019fine-tuning} and whether it can be effectively combined with adaptive temporal weighting within a unified pooling strategy.

\par
The main contributions of this paper are as follows:
\begin{itemize}
\item We provide the first systematic investigation of temporal pooling as an independent design variable in embedding-based training-free \ac{asd}, isolating its effect across multiple state-of-the-art embeddings and benchmark datasets.
\item We propose \ac{rdp}, a novel temporal pooling strategy that adaptively weights frames based on their temporal deviation from the sequence average.
\item Through extensive experiments on five benchmark datasets, we demonstrate that revisiting temporal pooling alone yields consistent and statistically significant performance gains, achieving state-of-the-art results for training-free \ac{asd} and surpassing previously reported trained systems on the DCASE2025 dataset.
\end{itemize}

\par
The remaining parts of this article are organized as follows.
In \cref{sec:related_work}, related literature for training-free anomalous sound detection with self-supervised audio embeddings is discussed.
In \cref{sec:methodology}, existing and novel temporal pooling approaches are presented.
The effectiveness of the temporal pooling strategies is experimentally evaluated in \cref{sec:results} using the setup from \cref{sec:setup} with several state-of-the-art embeddings on multiple datasets.
In addition, a few ablation studies are carried out.
The paper is concluded with a summary and possible extensions for future work in \cref{sec:conclusion}.

\section{Related Work}
\label{sec:related_work}
Temporal pooling is a fundamental component in many audio and speech processing tasks, where variable-length sequences must be aggregated into fixed-dimensional representations.
In speaker recognition, x-vector systems \cite{snyder2018x-vectors} aggregate frame-level features using simple statistics pooling, such as temporal mean and standard deviation, to obtain utterance-level embeddings that have been shown to be highly discriminative.
Similarly, in weakly labeled sound event detection, where only clip-level annotations are available, frame-level representations or predictions are commonly aggregated using simple pooling mechanisms, including mean or max pooling, to infer clip-level decisions \cite{kong2020sound}.
These approaches demonstrate that effective temporal aggregation can compensate for the absence of frame-level labels and highlight the importance of pooling strategies when operating on frame-level representations.

In the context of \ac{asd}, early studies have shown that relatively simple temporal pooling strategies can already achieve strong detection performance.
In particular, applying temporal mean or maximum pooling to \ac{stft}-based feature representations has been demonstrated to be effective for detecting anomalous machine operating sounds, even in the absence of explicit temporal modeling \cite{wilkinghoff2021sub-cluster}.
Subsequent work introduced \ac{gwrp} \cite{kolesnikov2016seed} as a temporal aggregation strategy for \ac{asd}, showing that it can outperform fixed pooling methods when applied to spectrogram-based features \cite{guan2023time-weighted}.
However, achieving these improvements requires optimizing the decay parameter separately for each machine type, which depends on access to machine-specific labels and implicitly leverages anomalous test data.
Existing embedding-based \ac{asd} approaches almost exclusively rely on temporal mean pooling to aggregate frame-level embeddings \cite{saengthong2024deep,saengthong2025retaining,wilkinghoff2025local,fujimura2025asdkit}.

Beyond training-free approaches, a related line of work incorporates learnable pooling mechanisms during fine-tuning. The AnoPatch framework \cite{jiang2024anopatch,jiang2025adaptive}, for instance, employs an attentive statistics pooling layer originally proposed for speaker recognition \cite{dawalatabad2021ecapa-tdnn}. Other works utilize a weighted mean with trainable weights \cite{han2024exploring}. While such approaches can improve performance, they rely on supervised or semi-supervised training, allowing the pooling mechanism to be optimized jointly with the embedding representation. In contrast, this work investigates temporal pooling for strictly training-free pipelines, where the embedding model remains fixed.

\section{Temporal pooling strategies}
\label{sec:methodology}
In this section, we discuss several temporal pooling strategies.
Let $\mathbf{X} = \{ \mathbf{x}_t \}_{t=1}^{T}$ denote a sequence of $T\in\mathbb{N}$ frame-level feature vectors extracted from an audio segment, where $\mathbf{x}_t\in\mathbb{R}^{D}$.
The goal of temporal pooling is to aggregate the sequence of feature vectors $\mathbf{X}$ into a single representation $\operatorname{Pool}(\mathbf{X})\in\mathbb{R}^D$ that can be used to distinguish between normal and anomalous samples.

\subsection{Mean pooling}
The most commonly used pooling strategy is to compute the temporal mean of the sequence, i.e.,
\BE \operatorname{MeanPool}(\mathbf{X}):=\frac{1}{T}\sum_{t=1}^T\mathbf{x}_t\in\mathbb{R}^D.\EE
Mean pooling averages the features over time, producing a representation that primarily reflects the typical characteristics of the recording. While averaging may help to suppress background noise and random fluctuations, it can also smooth out short or subtle anomalous events.

\subsection{Max pooling}
A complementary alternative to mean pooling is temporal max pooling, i.e., 
\BE
\operatorname{MaxPool}(\mathbf{X})
    :=
    \begin{pmatrix}
        \displaystyle \max_{1 \le t \le T} x_{t1} \\
        \displaystyle \max_{1 \le t \le T} x_{t2} \\
        \vdots \\
        \displaystyle \max_{1 \le t \le T} x_{tD}
    \end{pmatrix}
    \in \mathbb{R}^{D}.
\EE
In contrast to mean pooling, max pooling keeps only the strongest response observed over time, which may increase the influence of short, unusual events, but can also make the result more sensitive to random noise or brief spikes.

\subsection{\Acl{gwrp}}
\Ac{gwrp} \cite{kolesnikov2016seed} provides a smooth transition between mean and max pooling by weighting embeddings according to their strength, so that larger responses contribute more to the pooled representation while information from the entire sequence is retained.
Formally, \ac{gwrp} first rank-orders the values within each feature dimension of $\mathbf{X}$ in descending order. It is parameterized by a decay parameter $r\in[0,1]$ that controls the selectivity of the weighting and is defined as
\BE\operatorname{\ac{gwrp}}(\mathbf{X}; r):=\frac{1}{\sum_{t=1}^Tr^{t-1}}\sum_{t=1}^T
    \begin{pmatrix}
        x_{(t),1} \\
        x_{(t),2} \\
        \vdots \\
        x_{(t),D}
    \end{pmatrix}r^{t-1}\in\mathbb{R}^D\EE
where $x_{(t),j}$ denotes the $t$-th largest value of dimension $j$ in the sequence $\mathbf{X}$.
For $r=1$, this pooling strategy resembles mean pooling. For $r=0$, the convention $0^0=1$ is used in the weighting term, yielding max pooling.

\subsection{\Acl{gem} pooling}
\Ac{gem} pooling \cite{radenovic2019fine-tuning} is an alternative generalization of mean and max pooling defined as
\BE\operatorname{\acs{gem}Pool}(\mathbf{X}; p):=\bigg(\frac{1}{T}\sum_{t=1}^T\max\lbrace0,\mathbf{x}_t\rbrace^p\bigg)^{\frac{1}{p}}\in\mathbb{R}_{\geq 0}^D\EE
with $p\in\mathbb{R}_{>0}$.
All operations are applied element-wise across the embedding dimensions.
For $p=1$, this pooling strategy resembles mean pooling applied to non-negative entries.
The higher the parameter $p$, the more strongly large values dominate the aggregation, and in the limit $p\rightarrow\infty$, \ac{gem} converges to max pooling.
Note that negative entries are removed to preserve monotonicity for even integer values of $p$ and to avoid obtaining complex-valued embeddings for non-integer $p$.
We also experimented with taking the absolute value instead of setting all negative entries to zero, but this did not improve the performance.

\subsection{\Acl{rdp}}
As one of the main contributions of this paper, we propose \ac{rdp}, a training-free pooling method inspired by deviation pooling \cite{nafchi2015deviation}. Unlike classical deviation pooling, which computes relative deviations with respect to the entire feature sequence, \ac{rdp} uses these deviations to form a weighted temporal average where embeddings with larger deviations from the temporal average receive larger weights.
Such weighted aggregation is conceptually related to attention-based pooling strategies with learnable weights \cite{okabe2018attentive}, but in contrast, \ac{rdp} operates in a fully training-free manner.
\par
As a first step of \ac{rdp}, sample-wise deviations from the temporal mean are calculated as
\BE d_t:=\lVert \mathbf{x}_t-\operatorname{MeanPool}(\mathbf{X})\rVert_2\in\mathbb{R}_{\geq 0},\EE
and then normalized by
\BE\hat{d}_t:=\frac{d_t}{\max_{1\leq t'\leq T}d_{t'}}\in[0,1].\EE
Based on these deviations, weights that indicate how much individual embeddings deviate relatively to the entire sequence are computed as
\BE w^{\mathrm{RDP}}_t(\gamma):=\frac{(1+\hat{d}_t)^{\gamma}}{\sum_{t'=1}^T(1+\hat{d}_{t'})^{\gamma}}\in[0,1]\EE
with $\gamma\in\mathbb{R}_{\geq0}$.
Using these weights, the pooled representation is a weighted mean given by
\BE\operatorname{\ac{rdp}}(\mathbf{X}; \gamma):=\sum_{t=1}^Tw^{\mathrm{RDP}}_t(\gamma)\mathbf{x}_t\in\mathbb{R}^D.\EE
For $\gamma=0$, \ac{rdp} corresponds to mean pooling.    
The higher the value of $\gamma$, the more emphasis is placed on the embeddings strongly deviating from the mean.

\subsection{Hybrid extension of \acs{rdp}}

\Ac{rdp} and \ac{gem} operate on complementary aspects of the embedding sequence. Whereas \ac{rdp} assigns weights to entire embeddings based on their deviation from the temporal average, \ac{gem} performs feature-wise non-linear aggregation within each embedding dimension. This motivates combining adaptive temporal weighting with feature-wise non-linear aggregation. To this end, \ac{gem} pooling can be extended to a weighted formulation by introducing non-uniform positive weights $w_t \in \mathbb{R}_{>0}$ \cite{wang2022ranked},
\BE
\operatorname{W\ac{gem}Pool}(\mathbf{X}; w, p)
=
\left(
\frac{\sum_{t=1}^{T} w_t\, \max\{0,\mathbf{x}_t\}^{\,p}}
     {\sum_{t=1}^{T} w_t}
\right)^{1/p}
\in\mathbb{R}_{\geq 0}^D.
\EE
As in \ac{gem} pooling, all operations are applied element-wise to the embedding dimensions. In this work, we instantiate this weighted formulation using the weights derived from \ac{rdp}, i.e., by setting $w_t := w_t^{\mathrm{RDP}}(\gamma)$.

\subsection{Computational Complexity}
As temporal mean pooling is the standard aggregation strategy, the key question is how much additional computational overhead is introduced by alternative pooling methods. Mean pooling, max pooling, \ac{gem}, \ac{rdp}, and \ac{rdp}+\ac{gem} all scale linearly with the sequence length and embedding dimension, i.e., $\mathcal{O}(TD)$, whereas \ac{gwrp} requires sorting within each embedding dimension, resulting in a complexity of $\mathcal{O}(TD\log T)$. Consequently, all pooling strategies except \ac{gwrp} have the same asymptotic runtime complexity as mean pooling. All pooling strategies require $\mathcal{O}(D)$ additional memory for the pooled representation, while \ac{rdp} and \ac{rdp}+\ac{gem} additionally require $\mathcal{O}(T)$ storage for frame-wise weights and intermediate quantities. This overhead is dominated by the $\mathcal{O}(TD)$ memory required to store the embedding sequence itself.

\section{Experimental Setup}
\label{sec:setup}
\subsection{Datasets}
\begin{table*}[t]
	\centering
    \sisetup{
    detect-weight, 
    mode=text, 
    tight-spacing=true,
    round-mode=places,
    round-precision=0,
    table-format=3,
    table-number-alignment=center
}
	\caption{Overview of the considered DCASE \ac{asd} datasets, including recording durations and the number of recordings per section. Adapted from \cite{wilkinghoff2025local}.}
\begin{adjustbox}{max width=\textwidth}
	\begin{tabular}{lS[table-format=2]ccccccccccc}
		\toprule
        &&&&&&&&&\multicolumn{4}{c}{\# recordings (per section)}\\
        \cmidrule(lr){10-13}
        &\multicolumn{3}{c}{\# machine types}&\multicolumn{3}{c}{\# sections (per machine type)}&&&\multicolumn{2}{c}{source domain}&\multicolumn{2}{c}{target domain}\\
        \cmidrule(lr){2-4}\cmidrule(lr){5-7}\cmidrule(lr){10-11}\cmidrule(lr){12-13}
        Name&{total}&dev.\ set&eval.\ set&total&dev.\ set&eval.\ set& duration [\si{\second}]&split&{normal}&{anomalous}&{normal}&{anomalous}\\
		\midrule
		\multirow{2}{*}{DCASE2020 \cite{koizumi2020description}} & {\multirow{2}{*}{6}} & \multirow{2}{*}{6} & \multirow{2}{*}{6} & \multirow{2}{*}{6-7} & \multirow{2}{*}{3-4} & \multirow{2}{*}{3} & \multirow{2}{*}{10} & train & $\leq1000$ & 0 & 0 & 0\\
		&&&&&&&& test & $\leq400$ & $\leq200$ & 0 & 0\\
		\midrule
		\multirow{2}{*}{DCASE2022 \cite{dohi2022description}} & {\multirow{2}{*}{7}} & \multirow{2}{*}{7} & \multirow{2}{*}{7} & \multirow{2}{*}{6} & \multirow{2}{*}{3} & \multirow{2}{*}{3} & \multirow{2}{*}{10} & train & 990 & 0 & 10 & 0\\
		&&&&&&&& test & 50 & 50 & 50 & 50\\
		\midrule
		\multirow{2}{*}{DCASE2023 \cite{dohi2023description}} & {\multirow{2}{*}{14}} & \multirow{2}{*}{7} & \multirow{2}{*}{7} & \multirow{2}{*}{1} & \multirow{2}{*}{1} & \multirow{2}{*}{1} & \multirow{2}{*}{6--18} & train & 990 & 0 & 10 & 0\\
		&&&&&&&& test & 50 & 50 & 50 & 50\\
        \midrule
		\multirow{2}{*}{DCASE2024 \cite{nishida2024description}} & {\multirow{2}{*}{16}} & \multirow{2}{*}{7} & \multirow{2}{*}{9} & \multirow{2}{*}{1} & \multirow{2}{*}{1} & \multirow{2}{*}{1} & \multirow{2}{*}{6--18} & train & 990 & 0 & 10 & 0\\
		&&&&&&&& test & 50 & 50 & 50 & 50\\
		\midrule
		\multirow{2}{*}{DCASE2025 \cite{nishida2025description}} & {\multirow{2}{*}{14}} & \multirow{2}{*}{7} & \multirow{2}{*}{7} & \multirow{2}{*}{1} & \multirow{2}{*}{1} & \multirow{2}{*}{1} & \multirow{2}{*}{6--10} & train & 990 & 0 & 10 & 0\\
		&&&&&&&& test & 50 & 50 & 50 & 50\\
		\bottomrule
	\end{tabular}
\end{adjustbox}
\label{tab:datasets}
\end{table*}

We conduct experiments on five benchmark datasets from the DCASE challenge series, summarized in \Cref{tab:datasets}. These include the DCASE2020 \ac{asd} dataset \cite{koizumi2020description}, which is based on the MIMII corpus \cite{purohit2019mimii} and ToyADMOS \cite{koizumi2019toyadmos}; the DCASE2022 \ac{asd} dataset \cite{dohi2022description}, built from MIMII-DG \cite{dohi2022mimiidg} and ToyADMOS2 \cite{harada2021toyadmos2}; the DCASE2023 \ac{asd} dataset \cite{dohi2023description}, extending MIMII-DG with ToyADMOS2+ \cite{harada2023toyadmos2+}; the DCASE2024 \ac{asd} dataset \cite{nishida2024description}, which combines MIMII-DG, ToyADMOS2\# \cite{niizumi2024toyadmos2sharp}, and additional recordings collected using the IMAD-DS setup \cite{albertini2024imadds}; and the DCASE2025 \ac{asd} dataset \cite{nishida2025description}, consisting of MIMII-DG, ToyADMOS2025 \cite{harada2025toyadmos2025}, and further samples recorded under the same IMAD-DS conditions \cite{albertini2024imadds}.
\par
All datasets address semi-supervised acoustic anomaly detection for machine condition monitoring in realistic and noisy environments and include multiple machine types. Each dataset follows the official DCASE protocol and is divided into a development split and a separate evaluation split. In both splits, only recordings of normal operation are provided as reference data, while the corresponding test recordings contain both normal and anomalous samples. Except for DCASE2020, which contains data from a single domain, the datasets are designed to study domain generalization. To this end, the reference data consist of $990$ samples from a source domain and $10$ samples from a target domain. In the test data, machine types are known and the domain distributions are balanced, but explicit domain labels are not provided. No model parameters are learned from the reference data; they are solely used for distance-based anomaly scoring.
\par
The goal of the \ac{asd} system is to assign a continuous anomaly score to each test recording, where higher scores indicate a higher likelihood of abnormal behavior. All experiments follow the official evaluation protocols of the respective datasets. For DCASE2020, performance is measured using the arithmetic mean of the area under the ROC curve (AUC) and the partial AUC (pAUC) \cite{mcclish1989analyzing} with $p = 0.1$. For the remaining datasets, we report the metrics specified in their evaluation guidelines, namely the harmonic mean of the domain-specific AUCs and the domain-agnostic pAUC. Further details about the datasets are provided in the cited references.

\subsection{Audio Embedding Models}
\label{sec:emb_models}

A wide range of self-supervised audio embedding models has been proposed in recent years for various downstream audio tasks \cite{liu2022audio,yadav2025overview}.
In this work, we focus on four embedding models that are widely used and have demonstrated strong performance in \ac{asd}, namely OpenL3 \cite{cramer2019look}, BEATs \cite{chen2023beats}, \ac{eat} \cite{chen2024eat}, and Dasheng \cite{dinkel2024dasheng}.
These models represent different design philosophies and temporal resolutions, making them well suited for a comprehensive evaluation of temporal pooling strategies.
\par
Below, we describe the specific configurations and implementation details of the embedding models used in the experimental evaluation.
\par
\textbf{OpenL3}
We employ OpenL3 embeddings \cite{cramer2019look}, which are based on the Look, Listen and Learn framework \cite{arandjelovic2017look,arandjelovic2018objects}.
Input waveforms are segmented using a sliding window of \SI{1}{\second} duration with a hop size of \SI{0.1}{\second}, resulting in a temporal sequence of frame-level embeddings.
Performing temporal segmentation prior to pooling has been shown to yield improved performance compared to using a single clip-level embedding \cite{wilkinghoff2023using}.
For each segment, a 128-bin mel spectrogram is computed and passed through the OpenL3 model pre-trained on the environmental sound subset, producing embeddings of dimensionality $512$.
\par
\textbf{BEATs:}
BEATs \cite{chen2023beats} is one of the strongest and most widely adopted embedding models for ASD \cite{han2025exploring,fujimura2025asdkit}, achieving performance comparable to or exceeding that of \ac{asd}-specific foundation models such as ECHO \cite{zhang2025echo} and FISHER \cite{fan2025fisher}.
In this work, we use the official BEATs model pre-trained for three iterations on AudioSet \cite{gemmeke2017audioset}, without any additional fine-tuning.
All experiments are conducted using this frozen model.
\par
\begin{table}[!t]
\centering
\caption{Average performance obtained with EAT across the development and evaluation sets of the DCASE2020, DCASE2022, DCASE2023, DCASE2024, and DCASE2025 \acs{asd} datasets. $\Delta$ denotes improvement over baseline. CIs are 95\% paired-bootstrap intervals. Best performances in each column are in bold. All results are deterministic. }

\begin{adjustbox}{max width=\columnwidth,max totalheight=\textheight}
\begin{tabular}{cccccc}
\toprule
&&\multicolumn{2}{c}{\textbf{Mean Pooling}} &
\multicolumn{2}{c}{\textbf{Max Pooling}} \\
\cmidrule(lr){3-4} \cmidrule(lr){5-6}
min clamp & spike supp. & average & \scriptsize$\Delta$ (CI) 
& average & \scriptsize$\Delta$ (CI) \\
\midrule

& &
62.87\% & \color{gray}\scriptsize baseline &
63.33\% & \color{gray}\scriptsize baseline \\

&\checkmark&
64.35\% & \color{gray}\scriptsize +1.48 [1.22, 3.33] &
64.17\% & \color{gray}\scriptsize +0.83 [0.41, 1.35] \\


\checkmark&
&
65.58\% & \color{gray}\scriptsize +2.71 [2.35, 4.80] &
63.35\% & \color{gray}\scriptsize +0.02 [0.01, 0.04] \\


\checkmark&\checkmark&
\pmb{65.63\%} & \color{gray}\scriptsize +2.76 [2.19, 4.88] &
\pmb{64.30\%} & \color{gray}\scriptsize +0.97 [0.52, 1.48] \\


\bottomrule
\end{tabular}
\end{adjustbox}
\label{tab:prep}
\end{table}
\textbf{\ac{eat}:}
For \ac{eat} \cite{chen2024eat}, we use the official large-model checkpoint pre-trained for 20 epochs on AudioSet \cite{gemmeke2017audioset}.
We found that appropriate pre-processing of the extracted embeddings is crucial for obtaining competitive performance with \ac{eat}.
Specifically, hard thresholding of low-valued components (threshold set to 0.1) combined with suppression of large activation spikes using a hyperbolic tangent nonlinearity (applied to values above 0.5) proved essential.
These hyperparameters were optimized based on the performance on the development sets only.
The impact of this pre-processing step is quantitatively analyzed in \Cref{tab:prep} and constitutes an important practical insight for the use of \ac{eat} embeddings in \ac{asd}.
Note that we observed no measurable performance improvement for OpenL3, BEATs, or Dasheng when applying the same pre-processing. One possible explanation is that \ac{eat} embeddings exhibit higher dynamic range and less calibrated activation statistics than the other embedding models, which may make them more sensitive to low-magnitude components and extreme values. Under this interpretation, the pre-processing may help regularize the embedding distribution by suppressing low-valued components and limiting the influence of large activation spikes. However, a detailed analysis of the embedding statistics is beyond the scope of this work.

\par
\textbf{Dasheng:}
Dasheng \cite{dinkel2024dasheng} is a recently proposed audio foundation model designed for general-purpose audio representation learning.
In our experiments, we use the official Dasheng base model without additional fine-tuning.
\par
In addition to these models, we also evaluated several other self-supervised audio representations, including data2vec~2.0 \cite{baevski2022data2vec}, WavLM \cite{chen2022wavlm}, and Self-Supervised Audio Mamba (SSAM) \cite{yadav2024audio}.
Preliminary experiments indicated that applying more sophisticated pooling strategies also led to performance improvements over temporal mean pooling for these embeddings.
However, we refrain from including a detailed analysis in this work, as their overall performance as off-the-shelf representations for \ac{asd} remained substantially weaker and well below the state of the art.
A likely explanation is that these representations emphasize very short-term details, which may limit their ability to capture the longer-term acoustic context required to distinguish context-dependent anomalies from normal machine sounds.

\subsection{Attention-Based Pooling Baseline}

As a representative trainable pooling baseline, we additionally evaluated an attention-based pooling mechanism \cite{okabe2018attentive}. Frame-level embeddings are weighted according to attention scores predicted by a trainable scoring network, and the pooled representation is obtained as a weighted average of the embeddings. The attention network and a linear classification head are trained jointly on the official training split using an auxiliary classification task derived from the available machine metadata, where each unique combination of machine ID and attribute information is treated as a separate class and optimized using a cross-entropy loss. To improve generalization, mixup augmentation \cite{zhang2019mixup} is applied during training, where the interpolation coefficient is sampled from a uniform distribution on $[0,1]$. All models are trained for 50 epochs.

\par

Formally, the attention weights are given by
\[
a_t
=
\frac{\exp(f(\mathbf{x}_t))}
{\sum_{t'=1}^{T}\exp(f(\mathbf{x}_{t'}))}
\in [0,1],
\]
where $f:\mathbb{R}^{D}\rightarrow\mathbb{R}$ denotes a trainable scoring network. In our implementation, $f(\cdot)$ consists of a two-layer perceptron with a ReLU activation function and a hidden dimension of $128$. The pooled representation is then computed as
\[
\operatorname{AttPool}(\mathbf{X})
=
\sum_{t=1}^{T} a_t \mathbf{x}_t
\in\mathbb{R}^{D}.
\]

\par

Attention-based pooling is among the most widely used trainable aggregation mechanisms in audio and speech processing and has also been adopted in recent \ac{asd} systems such as AnoPatch~\cite{jiang2024anopatch,jiang2025adaptive}. Moreover, it provides a natural learned counterpart to \ac{rdp}, as both methods perform weighted temporal aggregation using frame-dependent weights. The key difference is that attention pooling learns these weights from metadata labels, whereas \ac{rdp} derives them directly from the embedding sequence in a fully training-free manner.

\subsection{Anomaly Score Calculation}
Let $\operatorname{Pool}:\mathbb{R}^{T\times D}\rightarrow \mathbb{R}^D$ denote a temporal pooling operator.
Further, let $\cX_\text{test}\subset \mathbb{R}^{T\times D}$ denote the set of test samples and $\cX_\text{ref}\subset \mathbb{R}^{T\times D}$ denote a reference set of normal training samples.
Then, anomaly scores are computed as the Euclidean distance between the temporally pooled embeddings of a test sample $\mathbf{X}\in \cX_\text{test}$ and its closest normal reference sample
\[
\begin{aligned}
\score(\mathbf{X},\mathcal{X}_{\text{ref}}\mid \operatorname{Pool})
&:= \min_{\mathbf{Y} \in \mathcal{X}_{\text{ref}}}
\pooldist(\mathbf{X},\mathbf{Y}\mid \operatorname{Pool})\\
&:= \min_{\mathbf{Y} \in \mathcal{X}_{\text{ref}}}
\left\lVert
\operatorname{Pool}(\mathbf{X})-\operatorname{Pool}(\mathbf{Y})
\right\rVert_2
\in \mathbb{R}_{\geq 0}.
\end{aligned}
\]
Following best practices, all available training samples are used as reference samples.
\par
To reduce performance degradations caused by domain shifts, we applied local density-based anomaly score normalization \cite{wilkinghoff2025keeping,wilkinghoff2025local} with $K=1$ and variance-minimization \cite{matsumoto2025adjusting} in log-space.
Formally, this corresponds to calculating anomaly scores of the form
\[
\begin{aligned}
\score_{\text{scaled}}(\mathbf{X},\mathcal{X}_{\text{ref}}\mid \operatorname{Pool},\alpha^*)
&:= \min_{\mathbf{Y}\in\mathcal{X}_{\text{ref}}}
\bigl(\log\pooldist(\mathbf{X},\mathbf{Y}\mid \operatorname{Pool}) \\
&\hphantom{:={}}\;-\alpha^*\log\pooldist(\mathbf{Y},\mathbf{Y}_{1}\mid \operatorname{Pool})\bigr)\in\mathbb{R}.
\end{aligned}
\]
where $\mathbf{Y}_{1}\neq \mathbf{Y}$ denotes the nearest neighbor in $\cX_\text{ref}$ to $\mathbf{Y}$, and
\[
\alpha^* \;=\;
\operatorname*{arg\,min}_{\alpha \in \mathbb{R}}
\operatorname{Var}_{\mathbf{Z} \sim \mathcal{X}_{\text{ref}}}
\!\left(
\score_{\text{scaled}}\!\left(\mathbf{Z}, \mathcal{X}_{\text{ref}} \mid\operatorname{Pool}, \alpha\right)
\right)\in\mathbb{R}.
\]
This particular scoring backend does not require any labels and does not make any assumptions about the data, which are important properties of a training-free \ac{asd} approach.
Moreover, since the normalization constants depend only on the reference samples, they can be pre-computed without introducing any additional computational overhead during inference.

\section{Results and Discussion}
\label{sec:results}
\subsection{Comparison of Pooling Strategies}
\begin{table*}[!t]
\centering
\caption{Average performance obtained with different pooling strategies and a representative trainable attention-pooling baseline across the development and evaluation sets of the DCASE2020, DCASE2022, DCASE2023, DCASE2024, and DCASE2025 \acs{asd} datasets. $\Delta$ denotes improvement over baseline. CIs are 95\% paired-bootstrap intervals. Highest numbers in each column are in bold. All results are based on the optimal hyperparameters determined on the development sets and are deterministic.}

\begin{adjustbox}{max width=\textwidth,max totalheight=\textheight}
\begin{NiceTabular}{c *{4}{cc}}
\toprule
& \multicolumn{8}{c}{\textbf{Embedding Model}} \\
\cmidrule{2-9}
\textbf{Pooling} 
& \multicolumn{2}{c}{OpenL3} 
& \multicolumn{2}{c}{BEATs} 
& \multicolumn{2}{c}{EAT} 
& \multicolumn{2}{c}{Dasheng} \\
\cmidrule(lr){2-3}
\cmidrule(lr){4-5}
\cmidrule(lr){6-7}
\cmidrule(lr){8-9}
& average & {\scriptsize$\Delta$(CI)}
& average & {\scriptsize$\Delta$(CI)}
& average & {\scriptsize$\Delta$(CI)}
& average & {\scriptsize$\Delta$(CI)}\\
\midrule
mean
& 64.65\% & {\color{gray}\scriptsize baseline}
& 67.01\% & {\color{gray}\scriptsize baseline}
& 65.63\% & {\color{gray}\scriptsize baseline}
& 63.10\% & {\color{gray}\scriptsize baseline}
\\

max
& 64.29\% & {\color{gray}\scriptsize -0.37 [-1.52, 0.76]}
& 68.12\% & {\color{gray}\scriptsize +1.11 [0.24, 1.90]}
& 64.30\% & {\color{gray}\scriptsize -1.33 [-2.37, -0.28]}
& 62.87\% & {\color{gray}\scriptsize -0.23 [-1.10, 0.67]}\\

\acs{gwrp}
& 65.34\% & {\color{gray}\scriptsize +0.69 [0.04, 1.37]}
& 68.38\% & {\color{gray}\scriptsize +1.38 [0.67, 2.06]}
& 65.61\% & {\color{gray}\scriptsize -0.01 [-0.59, 0.50]}
& 63.54\% & {\color{gray}\scriptsize +0.44 [-0.34, 1.16]}\\

\acs{gem}
& \pmb{65.40\%} & {\color{gray}\scriptsize +0.75 [0.32, 1.21]}
& 68.18\% & {\color{gray}\scriptsize +1.18 [0.56, 1.82]}
& \pmb{65.69\%} & {\color{gray}\scriptsize +0.07 [-0.11, 0.27]}
& 63.72\% & {\color{gray}\scriptsize +0.62 [0.02, 1.14]}\\

\acs{rdp}
& 64.80\% & {\color{gray}\scriptsize +0.15 [-0.62, 0.84]}
& \pmb{68.72\%} & {\color{gray}\scriptsize +1.71 [0.79, 2.70]}
& 65.62\% & {\color{gray}\scriptsize -0.00 [-0.06, 0.05]}
& \pmb{64.64\%} & {\color{gray}\scriptsize +1.53 [0.85, 2.27]}\\

\acs{rdp} + \acs{gem}
& 65.26\% & {\color{gray}\scriptsize +0.61 [0.11, 1.10]}
& 68.71\% & {\color{gray}\scriptsize +1.71 [0.87, 2.60]}
& \pmb{65.69\%} & {\color{gray}\scriptsize +0.06 [-0.14, 0.30]}
& 64.59\% & {\color{gray}\scriptsize +1.49 [0.67, 2.34]}\\

\midrule
attention
& 63.02\% & {\color{gray}\scriptsize -1.63 [-3.05, -0.27]}
& 67.03\% & {\color{gray}\scriptsize +0.02 [-2.12, 2.17]}
& 61.90\% & {\color{gray}\scriptsize -3.73 [-6.30, -1.34]}
& 62.02\% & {\color{gray}\scriptsize -1.08 [-3.34, 1.22]}\\

\bottomrule
\end{NiceTabular}
\end{adjustbox}
\label{tab:embs}
\end{table*}
\begin{table}[!t]
\centering
\caption{Best performing parameter settings across the development sets of the DCASE2020, DCASE2022, DCASE2023, DCASE2024, and DCASE2025 \acs{asd} datasets.}

\begin{adjustbox}{max width=\columnwidth,max totalheight=\textheight}
\begin{NiceTabular}{c *{5}{c}}
\toprule
\multirow{2}{*}{\textbf{Pooling}} & \multicolumn{4}{c}{\textbf{Embedding Model}} \\
\cmidrule{2-5}
& OpenL3 & BEATs & EAT & Dasheng & All\\
\midrule
mean&$-$&$-$&$-$&$-$&$-$\\
max&$-$&$-$&$-$&$-$&$-$\\
\acs{gwrp}&$r=0.9$&$r=0.4$&$r=0.9$&$r=0.6$&$r=0.7$\\
\acs{gem}&$p=9$&$p=10$&$p=3$&$p=6$&$p=6$\\
\acs{rdp}&$\gamma=10$&$\gamma=19$&$\gamma=1$&$\gamma=20$&$\gamma=10$\\
\acs{rdp} + \acs{gem} (p=3)&$\gamma=8$&$\gamma=16$&$\gamma=1$&$\gamma=20$&$\gamma=9$\\
\bottomrule
\end{NiceTabular}
\end{adjustbox}
\label{tab:params}
\end{table}

As an initial experiment, we compare the temporal pooling strategies introduced in \Cref{sec:methodology} across different audio embedding models (see \Cref{sec:emb_models}). The results are reported in \Cref{tab:embs}. For all experiments, hyperparameters were fixed to the values specified in \Cref{tab:params}. These values were selected using the labeled test sets of the official development splits and the corresponding evaluation metrics. No information from the evaluation splits was used. The selected parameters are fixed globally for each embedding model and are not tuned separately for individual datasets. Their sensitivity is analyzed in \Cref{sec:hyper_params}. Since temporal mean pooling constitutes the de facto standard aggregation mechanism in embedding-based training-free \ac{asd}, it serves as the primary baseline throughout our analysis.
\par
The results indicate that both the optimal choice of a pooling strategy and its effectiveness are strongly dependent on the underlying embedding model.
A consistent trend emerges when comparing simple pooling baselines to more advanced methods: When the performance of maximum pooling is comparable to or exceeds mean pooling, advanced pooling strategies tend to yield the largest performance improvements.
Conversely, for embeddings such as \ac{eat}, where maximum pooling performs considerably worse than mean pooling, none of the advanced pooling approaches provides statistically significant improvements.
A plausible explanation for the limited gains observed with advanced temporal pooling for \ac{eat} lies in the pre-processing applied to the embeddings.
After normalization, \ac{eat} embeddings exhibit reduced temporal variance and a more uniform distribution of anomaly-relevant information across time.
Consequently, mean pooling is already near-optimal, and more sophisticated pooling strategies, which primarily exploit temporal sparsity or extreme activations, offer little additional benefit.
In contrast, OpenL3, BEATs, and Dasheng embeddings show greater variation over time, allowing advanced pooling methods to improve performance.
\par
\begin{table}[!t]
\centering
\caption{Average performance obtained with different pooling strategies for the embedding models OpenL3, BEATs, EAT, and Dasheng. Results are averages across the development and evaluation sets of the DCASE2020, DCASE2022, DCASE2023, DCASE2024, and DCASE2025 \acs{asd} datasets. $\Delta$ denotes improvement over baseline. CIs are 95\% paired-bootstrap intervals. Highest numbers in each column are in bold. All results are based on the optimal hyperparameters determined on the development sets and are deterministic.}

\begin{adjustbox}{max width=\textwidth,max totalheight=\textheight}
\begin{NiceTabular}{c *{2}{cc}}
\toprule
& \multicolumn{4}{c}{\textbf{Hyperparameter Settings}} \\
\cmidrule(lr){2-5}
\textbf{Pooling}
& \multicolumn{2}{c}{embedding-agnostic} 
& \multicolumn{2}{c}{embedding-specific} \\
\cmidrule(lr){2-3}
\cmidrule(lr){4-5}
& average & {\scriptsize$\Delta$(CI)}
& average & {\scriptsize$\Delta$(CI)} \\
\midrule
mean
& 65.10\% & {\color{gray}\scriptsize baseline}
& 65.10\% & {\color{gray}\scriptsize baseline} \\

max
& 64.89\% & {\color{gray}\scriptsize -0.21 [-0.77, 0.36]}
& 64.89\% & {\color{gray}\scriptsize -0.21 [-0.77, 0.36]}
\\

\acs{gwrp}
& 65.48\% & {\color{gray}\scriptsize +0.38 [-0.03, 0.79]}
& 65.72\% & {\color{gray}\scriptsize +0.62 [0.26, 0.99]}
\\

\acs{gem}
& 65.73\% & {\color{gray}\scriptsize +0.63 [0.35, 0.92]}
& 65.75\% & {\color{gray}\scriptsize +0.65 [0.38, 0.93]}
\\

\acs{rdp}
& 65.50\% & {\color{gray}\scriptsize +0.40 [0.03, 0.75]}
& 65.94\% & {\color{gray}\scriptsize +0.84 [0.43, 1.28]}
\\

\acs{rdp} + \acs{gem}
& \pmb{65.76\%} & {\color{gray}\scriptsize +0.66 [0.24, 1.07]}
& \pmb{66.06\%} & {\color{gray}\scriptsize +0.96 [0.59, 1.37]}
\\

\bottomrule
\end{NiceTabular}
\end{adjustbox}
\label{tab:avg_perf}
\end{table}
The effectiveness of pooling strategies varies across embeddings. \Ac{rdp} yields the highest average performance for BEATs and Dasheng embeddings. In contrast, \ac{rdp} is less effective for OpenL3, where \ac{gem} pooling performs best. Similarly, \ac{gem} pooling achieves the highest average performance for \ac{eat} embeddings. \Ac{gwrp} improves the average performance over mean pooling but generally achieves lower performance than \ac{gem} and \ac{rdp}. Across all embeddings, the hybrid \ac{rdp}+\ac{gem} variant does not consistently outperform either \ac{gem} or \ac{rdp} individually, but achieves average performance comparable to the strongest individual pooling strategy for each embedding model.

\par

The attention-based pooling baseline yielded mixed results across embeddings and did not provide consistent improvements over temporal mean pooling. While attention pooling occasionally yielded substantial improvements for individual embedding--dataset combinations (e.g., increasing the development/evaluation scores from 64.22/63.25 to 69.39/65.33 on DCASE2022 using BEATs embeddings), it also degraded performance in other cases (e.g., from 58.40/60.28 to 59.24/54.92 on DCASE2024 using the same embeddings). Consequently, attention pooling underperformed all training-free pooling strategies on average. This may be because optimizing temporal weights using metadata-defined classification objectives encourages the model to emphasize metadata-discriminative rather than anomaly-discriminative temporal regions. For example, in DCASE2024, machine-specific background noise may correlate with the metadata labels while carrying no information about anomalies, leading the attention mechanism to focus on these background characteristics instead of anomaly-relevant acoustic events. In contrast, the proposed training-free pooling strategies yielded statistically significant performance improvements across embeddings and datasets.

\par

\begin{figure*}
    \centering
    \begin{adjustbox}{max width=\textwidth}
          \input{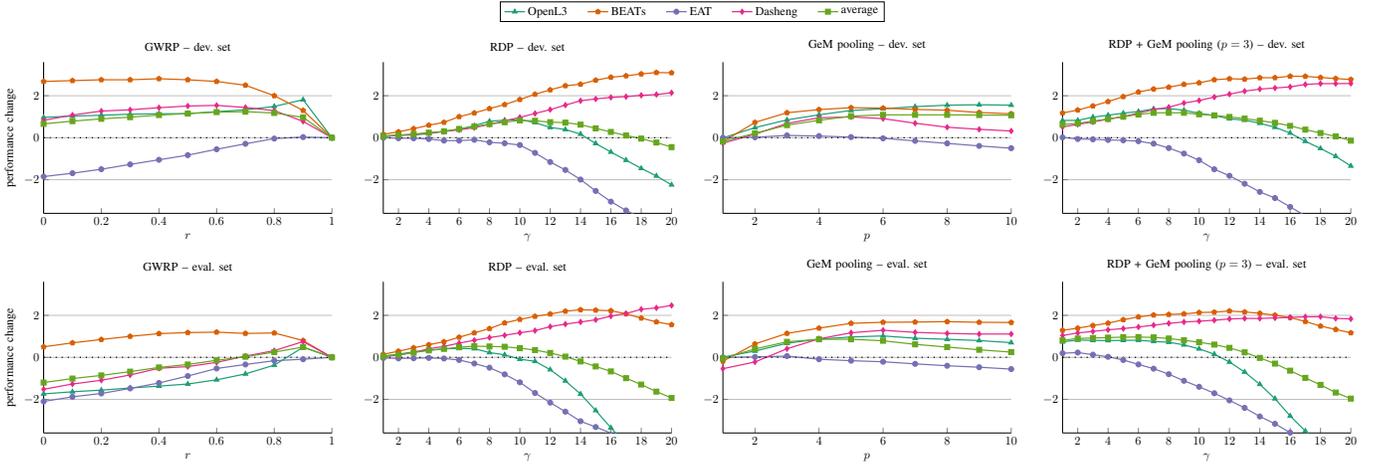}
    \end{adjustbox}
    \caption{Sensitivity analysis of the performance with respect to the hyperparameters of different pooling strategies. Relative performance changes (\%) compared to mean pooling are depicted. All values are geometric means across the development and evaluation sets of the DCASE2020, DCASE2022, DCASE2023, DCASE2024, and DCASE2025 datasets.}
    \label{fig:scaling_ablation}
\end{figure*}
\par
To assess whether the observed improvements depend on embedding-specific tuning, we compare embedding-agnostic and embedding-specific hyperparameter settings, as summarized in \Cref{tab:avg_perf}. Overall, max pooling performs slightly worse than temporal mean pooling, although the difference is not statistically significant, while \ac{gwrp} provides moderate improvements that lose statistical significance under embedding-agnostic settings. In contrast, both \ac{gem} and \ac{rdp} yield consistent and statistically significant gains over mean pooling even without embedding-specific tuning, indicating that the benefits of advanced temporal aggregation are not merely a consequence of hyperparameter adaptation. Among these methods, \ac{gem} exhibits the most stable performance across both regimes, whereas \ac{rdp} improves the average performance by 0.84 percentage points under embedding-specific hyperparameter settings compared with 0.40 percentage points under embedding-agnostic settings, suggesting that it benefits from adaptation to the characteristics of individual embedding models. Across all strategies, the hybrid \ac{rdp}+\ac{gem} variant achieves average performance comparable to the strongest individual pooling strategies, indicating that combining adaptive temporal weighting and feature-wise non-linear aggregation does not compromise performance across the evaluated embedding models.

\par
Although the absolute improvements are modest, they are obtained solely by replacing the temporal pooling operation without modifying the embedding model or anomaly scoring backend. Moreover, the improvements achieved by \ac{gem}, \ac{rdp}, and \ac{rdp}+\ac{gem} remain statistically significant across datasets and embeddings. These findings demonstrate that temporal pooling constitutes an important yet previously overlooked design choice in training-free ASD. Rather than adopting temporal mean pooling by default, the pooling strategy should be selected according to the characteristics of the embedding representation.

\subsection{Hyperparameter Sensitivity Analysis}
\label{sec:hyper_params}

\begin{table*}
\centering
\caption{Official performance metrics and dataset-wise harmonic means for different \acs{asd} systems. All proposed systems employ the hybrid \acs{rdp}+\ac{gem} pooling strategy with embedding-specific, dataset-independent parameter settings. All values are reported in percentages. The best result in each column is shown in bold, while the best training-free result is underlined.}
\begin{threeparttable}
\begin{adjustbox}{max width=\textwidth}
\begin{NiceTabular}{lcc*{14}{c}} 
\toprule
&&\multicolumn{3}{c}{DCASE2020 \cite{koizumi2020description}}&\multicolumn{3}{c}{DCASE2022 \cite{dohi2022description}}&\multicolumn{3}{c}{DCASE2023 \cite{dohi2023description}}&\multicolumn{3}{c}{DCASE2024 \cite{nishida2024description}}&\multicolumn{3}{c}{DCASE2025 \cite{nishida2025description}}\\
\cmidrule(lr){3-5}\cmidrule(lr){6-8}\cmidrule(lr){9-11}\cmidrule(lr){12-14}\cmidrule(lr){15-17}
\acs{asd} system&training-free&
dev.&eval.&mean&dev.&eval.&mean&dev.&eval.&mean&dev.&eval.&mean&dev.&eval.&mean\\
\midrule
openL3&\checkmark&$80.5$&$81.8$&$81.2$&$61.9$&$64.0$&$62.9$&$60.3$&$66.2$&$63.1$&$57.9$&$60.8$&$59.3$&$60.6$&$58.6$&$59.6$\\
BEATs&\checkmark&$\underline{86.8}$&$\underline{87.4}$&$\underline{87.1}$&$\underline{66.1}$&$\underline{65.1}$&$\underline{65.6}$&$\underline{65.7}$&$70.8$&$\underline{68.2}$&$\underline{58.7}$&$60.1$&$59.4$&$\underline{64.0}$&$\pmb{\underline{62.4}}$&$\pmb{\underline{63.2}}$\\
EAT&\checkmark&$79.1$&$82.2$&$80.6$&$63.2$&$62.5$&$62.8$&$62.9$&$65.6$&$64.2$&$58.9$&$60.9$&$59.9$&$63.0$&$58.5$&$60.7$\\
Dasheng&\checkmark&$80.9$&$81.6$&$81.2$&$60.5$&$61.8$&$61.1$&$62.4$&$64.0$&$63.2$&$57.1$&$58.5$&$57.8$&$61.3$&$57.9$&$59.6$\\
\midrule
Wilkinghoff (\ac{stft}) \cite{wilkinghoff2021sub-cluster} &\checkmark&$73.6$&$75.4$&$74.5$&$-$&$-$&$-$&$-$&$-$&$-$&$-$&$-$&$-$&$-$&$-$&$-$\\
Saengthong et al. (BEATs) \cite{saengthong2024deep}&\checkmark&$74.7$\tnote{a}&$-$&$-$&$-$&$-$&$-$&$-$&$\underline{73.8}$\tnote{a}&$-$&$-$&$-$&$-$&$-$&$-$&$-$\\
Fujimura et al. (BEATs) \cite{fujimura2025asdkit}&\checkmark&$76.9$&$77.5$&$77.2$&$61.5$&$60.3$&$60.9$&$62.3$&$62.6$&$62.4$&$57.7$&$55.7$&$56.7$&$-$&$-$&$-$\\
Wilkinghoff et al. (BEATs) \cite{wilkinghoff2025local}&\checkmark&$81.5$&$82.2$&$81.8$&$-$&$-$&$-$&$64.8$&$67.6$&$66.2$&$58.1$&$\underline{62.4}$&$\underline{60.2}$&$-$&$-$&$-$\\
Fan et al. (FISHER) \cite{fan2025fisher}&\checkmark&$-$&$-$&$71.0$&$-$&$-$&$59.6$&$-$&$-$&$62.6$&$-$&$-$&$55.6$&$-$&$-$&$-$\\
Zhang et al. (ECHO) \cite{zhang2025echo}&\checkmark&$-$&$-$&$72.2$&$-$&$-$&$60.0$&$-$&$-$&$63.7$&$-$&$-$&$57.9$&$-$&$-$&$58.7$\\
\midrule
Koizumi et al. \cite{koizumi2020description}&$-$&$66.6$&$70.0$&$68.3$&$-$&$-$&$-$&$-$&$-$&$-$&$-$&$-$&$-$&$-$&$-$&$-$\\
Wilkinghoff \cite{wilkinghoff2021sub-cluster}&$-$&$90.7$&$92.8$&$91.7$&$-$&$-$&$-$&$-$&$-$&$-$&$-$&$-$&$-$&$-$&$-$&$-$\\
Liu et al. \cite{liu2022anomalous}&$-$&$89.4$&$-$&$-$&$-$&$-$&$-$&$-$&$-$&$-$&$-$&$-$&$-$&$-$&$-$&$-$\\
Harada et al. \cite{harada2023first}&$-$&$-$&$-$&$-$&$-$&$59.0$&$-$&$-$&$61.1$&$-$&$-$&$56.5$&$-$&$-$&$56.5$&$-$\\
Wilkinghoff \cite{wilkinghoff2023design}&$-$&$-$&$-$&$-$&$62.8$&$63.0$&$62.9$&$-$&$-$&$-$&$-$&$-$&$-$&$-$&$-$&$-$\\
Hou et al. \cite{hou2023decoupling}&$-$&$88.8$&$92.0$&$90.4$&$-$&$-$&$-$&$-$&$-$&$-$&$-$&$-$&$-$&$-$&$-$&$-$\\
Wilkinghoff \cite{wilkinghoff2024self}&$-$&$-$&$-$&$-$&$-$&$-$&$-$&$64.2$&$66.6$&$65.4$&$-$&$-$&$-$&$-$&$-$&$-$\\
Han et al. \cite{han2024exploring}&$-$&$-$&$-$&$-$&$-$&$-$&$-$&$64.3$&$-$&$-$&$-$&$-$&$-$&$-$&$-$&$-$\\
Zhang et al. \cite{zhang2024dual}&$-$&$-$&$-$&$-$&$-$&$-$&$-$&$-$&$71.3$&$-$&$-$&$-$&$-$&$-$&$-$&$-$\\
Jiang et al. \cite{jiang2024anopatch,jiang2025adaptive}&$-$&$90.9$&$\pmb{94.3}$&$92.6$&$-$&$-$&$-$&$64.2$&$\pmb{74.2}$&$68.8$&$62.5$&$65.6$&$64.0$&$-$&$-$&$-$\\
Wilkinghoff \cite{wilkinghoff2024adaproj}&$-$&$-$&$-$&$-$&$73.1$&$67.1$&$70.0$&$62.9$&$64.5$&$63.7$&$-$&$-$&$-$&$-$&$-$&$-$\\
Yin et al. \cite{yin2024self}&$-$&$-$&$-$&$-$&$-$&$-$&$-$&$68.1$&$-$&$-$&$-$&$-$&$-$&$-$&$-$&$-$\\
Fujimura et al. \cite{fujimura2025improvements}&$-$&$-$&$-$&$-$&$-$&$-$&$-$&$67.2$&$68.8$&$68.0$&$67.6$&$62.0$&$64.7$&$-$&$-$&$-$\\
Yin et al. \cite{yin2025diffusion}&$-$&$-$&$-$&$-$&$-$&$-$&$-$&$-$&$-$&$-$&$-$&$\pmb{67.1}$&$-$&$-$&$-$&$-$\\
Jiang et al. \cite{jiang2025adaptive}&$-$&$-$&$-$&$-$&$-$&$-$&$-$&$-$&$-$&$-$&$64.1$&$66.0$&$65.0$&$-$&$-$&$-$\\
Fujimura et al. \cite{fujimura2025asdkit}&$-$&$90.4$&$93.5$&$91.9$&$\pmb{73.9}$&$69.9$&$\pmb{71.8}$&$64.0$&$72.0$&$67.8$&$59.9$&$61.5$&$60.7$&$-$&$-$&$-$\\
Matsumoto et al. \cite{matsumoto2025adjusting}&$-$&$-$&$-$&$-$&$71.0$&$68.8$&$69.9$&$66.5$&$69.0$&$67.7$&$62.7$&$57.1$&$59.8$&$-$&$-$&$-$\\
Wilkinghoff et al. \cite{wilkinghoff2025local}&$-$&$\pmb{94.2}$&$93.3$&$\pmb{93.7}$&$-$&$-$&$-$&$\pmb{71.3}$&$72.4$&$\pmb{71.8}$&$65.2$&$56.5$&$60.5$&$-$&$-$&$-$\\
Fujimura et al. \cite{fujimura2025discriminative}&$-$&$-$&$-$&$-$&$-$&$-$&$-$&$-$&$-$&$-$&$-$&$-$&$-$&$\pmb{64.9}$&$60.0$&$62.4$\\
Han et al. \cite{han2025exploring}&$-$&$91.1$&$93.1$&$92.1$&$72.4$&$66.6$&$69.4$&$64.3$&$68.7$&$66.4$&$64.1$&$65.5$&$64.8$&$-$&$-$&$-$\\
\midrule
DCASE2020 Challenge winner \cite{giri2020unsupervised}&$-$&$-$&$89.8$&$-$&$-$&$-$&$-$&$-$&$-$&$-$&$-$&$-$&$-$&$-$&$-$&$-$\\
DCASE2022 Challenge winner \cite{zeng2022robust}&$-$&$-$&$-$&$-$&$-$&$\pmb{71.0}$&$-$&$-$&$-$&$-$&$-$&$-$&$-$&$-$&$-$&$-$\\
DCASE2023 Challenge winner \cite{jie2023anomalous}&$-$&$-$&$-$&$-$&$-$&$-$&$-$&$-$&$67.0$&$-$&$-$&$-$&$-$&$-$&$-$&$-$\\
DCASE2024 Challenge winner \cite{lv2024aithu}&$-$&$-$&$-$&$-$&$-$&$-$&$-$&$-$&$-$&$-$&$\pmb{67.8}$&$66.2$&$\pmb{67.0}$&$-$&$-$&$-$\\
DCASE2025 Challenge winner \cite{wang2025pre-trained}&$-$&$-$&$-$&$-$&$-$&$-$&$-$&$-$&$-$&$-$&$-$&$-$&$-$&$60.9$&$61.6$&$61.2$\\
\bottomrule
\end{NiceTabular}
\end{adjustbox}
\begin{tablenotes}\footnotesize
\item [a] Obtained by using a domain-wise standardization of the test scores, which requires domain labels and destroys independence between test samples.
\end{tablenotes}
\end{threeparttable}
\label{tab:sota}
\end{table*}

Next, we analyze the performance sensitivity with respect to the pooling hyperparameters and justify the choice of the embedding-specific parameter values.
The corresponding results are shown in \Cref{fig:scaling_ablation}.
Overall, with the exception of \ac{gem} pooling, most pooling strategies are sensitive to the chosen hyperparameters.
Moreover, the results suggest that the resulting performance improvements appear to be more strongly influenced by the choice of embedding model than by the dataset.
This behavior is particularly evident for \ac{gwrp}, for which a sharp performance peak is observed around $r=0.9$ for OpenL3 embeddings on the development set and for several embedding models on the evaluation set, while such a peak is absent for others, most notably BEATs.
At the same time, for the other pooling approaches, the overall trends and relative ordering of performance remain similar between the aggregated development and evaluation results, suggesting that the observed improvements are not dominated by overfitting to the development data.
This consistency suggests that the selected hyperparameters generalize well beyond the development data. Although the proposed methods introduce additional hyperparameters, these are selected once per embedding model and reused across datasets, limiting the tuning effort while still providing consistent performance improvements.

\subsection{Comparison to the State-of-the-Art}

Having established that improved pooling yields consistent gains across embeddings and hyperparameter settings, we finally compare the proposed approach with a broad range of previously reported \ac{asd} systems. These include both training-free approaches and methods that exploit metadata or auxiliary labels through task-specific training.
For completeness, we also include the winning systems of the corresponding DCASE challenges for each dataset.
As a result, the comparison is intended to provide a broad performance reference rather than a strictly controlled benchmark.
The comparative results are summarized in \Cref{tab:sota}.
Across the evaluated datasets, the choice of embedding model strongly influences overall performance.
In particular, BEATs yields substantially higher scores than the other embedding models on most datasets.
On the DCASE2024 dataset, however, BEATs performs comparably to OpenL3 and slightly below EAT, underscoring the importance of representation quality for training-free \ac{asd}.
\par
On average, systems incorporating the proposed temporal pooling strategies outperform previously reported training-free methods across the majority of evaluated datasets, despite not relying on domain labels, machine-specific tuning, or additional constraints on the evaluation protocol.
Although methods that exploit metadata or auxiliary labels through task-specific training generally achieve stronger performance, the combination of improved temporal pooling and the normalization approach reduces the performance difference to these methods. In several cases, the proposed method matches or exceeds previously reported trained systems, including challenge-winning approaches on the DCASE2023 and DCASE2025 datasets.
The remaining performance difference on the DCASE2023 evaluation set is largely attributable to domain-wise score standardization employed in \cite{saengthong2024deep}, which assumes access to domain labels and introduces dependencies between test samples that are not permitted under the strictly training-free evaluation protocol adopted here. Notably, on the DCASE2025 dataset, the proposed approach achieves the highest performance among the methods included in \Cref{tab:sota}, surpassing previously reported trained systems and ensemble-based approaches.
One plausible explanation is that DCASE2025 contains machine-specific background noise conditions. Such conditions may become predictive of machine identity and therefore be exploited by trained systems. As a result, the learned representations may rely on background characteristics rather than the target machine sounds, reducing their sensitivity to anomalies. In contrast, training-free approaches do not rely on metadata-defined discrimination and are therefore not exposed to this particular failure mode.
\par
These findings demonstrate that temporal pooling should be treated as a tunable design choice rather than a fixed component in training-free \ac{asd}.
Importantly, all reported results are obtained without supervised training, test-set adaptation, or domain-wise standardization.

\section{Conclusion}
\label{sec:conclusion}

This paper presented the first systematic study of temporal pooling in embedding-based training-free \ac{asd}, a component that has largely remained fixed to simple mean aggregation.
Through experiments across four state-of-the-art embedding models and five benchmark datasets, we demonstrated that temporal pooling should be treated as a tunable design choice rather than a fixed component of embedding-based \ac{asd} pipelines.
We introduced \ac{rdp}, a novel training-free pooling strategy, and further investigated a hybrid extension combining \ac{rdp} with \ac{gem} pooling.
Together, the proposed pooling strategies achieved state-of-the-art performance for training-free \ac{asd} and, on DCASE2025, surpassed previously reported trained systems and ensembles.

The observed gains are comparable in magnitude to those obtained by switching between commonly used embedding models, illustrating that improvements in embedding-based anomaly detection do not necessarily require larger or more sophisticated foundation models.
Instead, carefully revisiting overlooked components such as temporal aggregation can yield statistically significant performance gains at essentially no additional training cost.
Future work will investigate which properties of pretrained embeddings determine the effectiveness of different temporal pooling strategies, enabling more principled selection of pooling methods for future foundation models.
Overall, our results demonstrate that temporal aggregation is a critical design choice in embedding-based anomaly detection systems and should be considered alongside embedding selection when designing future training-free \ac{asd} pipelines.

\section{Generative AI disclosure}
Generative AI tools were used for language editing and polishing of the manuscript.
All scientific content, interpretations, and conclusions are the responsibility of the authors.

\bibliographystyle{IEEEtran}
\bibliography{refs}

@string{icassp = "Proc. ICASSP"}

@string{interspeech = "Proc. Interspeech"}

@string{waspaa = "Proc. WASPAA"}

@string{dcase = "Proc. DCASE"}

@string{eusipco = "Proc. EUSIPCO"}

@string{apsipa = "Proc. APSIPA"}

@string{ieee-tpami = "IEEE Trans. Pattern Anal. Mach. Intell."}

@string{ieee-taslp = "IEEE Trans. Audio, Speech, Lang. Process."}

@string{ieee-acm-taslp = "IEEE/ACM Trans. Audio, Speech, Lang. Process."}

@string{icml = "Proc. ICML"}

@string{iclr = "Proc. ICLR"}

@string{eccv = "Proc. ECCV"}

@string{iccv = "Proc. ICCV"}

@string{ijcnn = "Proc. IJCNN"}

@string{ijcai = "Proc. IJCAI"}

@inproceedings{dohi2022description,
  author       = {Kota Dohi and others},
  title        = {Description and Discussion on {DCASE} 2022 {Challenge} {Task} 2: {U}nsupervised Anomalous Sound Detection for Machine Condition Monitoring Applying Domain Generalization Techniques},
  booktitle    = dcase,
  year         = {2022}
}

@inproceedings{dohi2023description,
  author       = {Kota Dohi and others},
  title  = {Description and Discussion on {DCASE} 2023 {Challenge} {Task} 2: {F}irst-Shot Unsupervised Anomalous Sound Detection for Machine Condition Monitoring},
  booktitle = dcase,
  year   = {2023}
}

@inproceedings{nishida2025description,
  author       = {Nishida, Tomoya and others},
  title        = {Description and Discussion on {DCASE} 2025 Challenge Task 2: First-shot Unsupervised Anomalous Sound Detection for Machine Condition Monitoring},
  booktitle    = dcase,
  year         = {2025}
}

@inproceedings{koizumi2020description,
  author       = {Yuma Koizumi and others},
  title        = {Description and Discussion on {DCASE2020} {Challenge} {Task2}: {U}nsupervised Anomalous Sound Detection for Machine Condition Monitoring},
  booktitle    = dcase,
  year         = {2020}
}

@inproceedings{nishida2024description,
  author       = {Tomoya Nishida and others},
  title        = {Description and Discussion on {DCASE} 2024 {Challenge} {Task} 2: {F}irst-Shot Unsupervised Anomalous Sound Detection for Machine Condition Monitoring},
  booktitle    = dcase,
  year         = {2024},
}

@inproceedings{jiang2024anopatch,
  title     = {{A}no{P}atch: {T}owards Better Consistency in Machine Anomalous Sound Detection},
  author    = {Anbai Jiang and others},
  year      = {2024},
  booktitle = interspeech,
}

@inproceedings{saengthong2024deep,
   title={Deep Generic Representations for Domain-Generalized Anomalous Sound Detection}, 
   author={Phurich Saengthong and Takahiro Shinozaki},
   booktitle = icassp,
   year = {2025},
}

@inproceedings{chen2023beats,
  author       = {Sanyuan Chen and others},
  title        = {{BEATs}: {A}udio Pre-Training with Acoustic Tokenizers},
  booktitle    = icml,
  year         = {2023}
}

@inproceedings{chen2024eat,
  author       = {Wenxi Chen and
                  Yuzhe Liang and
                  Ziyang Ma and
                  Zhisheng Zheng and
                  Xie Chen},
  title        = {{EAT:} Self-Supervised Pre-Training with Efficient Audio Transformer},
  booktitle    = ijcai,
  year         = {2024},
}

@inproceedings{matsumoto2025adjusting,
    author = "Matsumoto, Masaaki and Fujimura, Takuya and Huang, WenChin and Toda, Tomoki",
    title = "Adjusting Bias in Anomaly Scores via Variance Minimization for Domain-Generalized Discriminative Anomalous Sound Detection",
    booktitle = dcase,
    year = "2025",
}

@inproceedings{fujimura2025improvements,
  author       = {Takuya Fujimura and
                  Ibuki Kuroyanagi and
                  Tomoki Toda},
  title        = {Improvements of Discriminative Feature Space Training for Anomalous Sound Detection in Unlabeled Conditions},
  booktitle    = icassp,
  year         = {2025}
}

@inproceedings{wilkinghoff2025keeping,
  author       = {Kevin Wilkinghoff and
                  Haici Yang and
                  Janek Ebbers and
                  Fran{\c{c}}ois G. Germain and
                  Gordon Wichern and
                  Jonathan Le Roux},
  title        = {Keeping the Balance: Anomaly Score Calculation for Domain Generalization},
  booktitle    = icassp,
  year         = {2025}
}

@article{wilkinghoff2025local,
  author       = {Kevin Wilkinghoff and
                  Haici Yang and
                  Janek Ebbers and
                  Fran{\c{c}}ois G. Germain and
                  Gordon Wichern and
                  Jonathan Le Roux},
  title        = {Local Density-Based Anomaly Score Normalization for Domain Generalization},
  journal      = ieee-taslp,
  year         = {2025},
  volume       = {33}
}

@inproceedings{wilkinghoff2021sub-cluster,
  author       = {Kevin Wilkinghoff},
  title        = {Sub-Cluster {AdaCos}: Learning Representations for Anomalous Sound Detection},
  booktitle    = ijcnn,
  year         = {2021}
}

@inproceedings{fujimura2025asdkit,
    author = "Fujimura, Takuya and Wilkinghoff, Kevin and Imoto, Keisuke and Toda, Tomoki",
    title = "{ASDKit}: A Toolkit for Comprehensive Evaluation of Anomalous Sound Detection Methods",
    booktitle = dcase,
    year = "2025",
}

@article{wilkinghoff2024why,
  author       = {Kevin Wilkinghoff and
                  Frank Kurth},
  title        = {Why Do Angular Margin Losses Work Well for Semi-Supervised Anomalous Sound Detection?},
  journal      = ieee-acm-taslp,
  year         = {2024},
  volume       = {32}
}

@inproceedings{fujimura2025discriminative,
    author = "Fujimura, Takuya and Kuroyanagi, Ibuki and Toda, Tomoki",
    title = "Discriminative Anomalous Sound Detection Using Pseudo Labels, Target Signal Enhancement, and Ensemble Feature Extractors",
    booktitle = dcase,
    year = "2025",
}

@inproceedings{cramer2019look,
  author    = {Aurora Cramer and
               Ho{-}Hsiang Wu and
               Justin Salamon and
               Juan Pablo Bello},
  title     = {Look, Listen, and Learn More: Design Choices for Deep Audio Embeddings},
  booktitle = icassp,
  year      = {2019}
}

@inproceedings{wilkinghoff2025handling,
    author = "Wilkinghoff, Kevin and Fujimura, Takuya and Imoto, Keisuke and Le Roux, Jonathan and Tan, Zheng-Hua and Toda, Tomoki",
    title = "Handling Domain Shifts for Anomalous Sound Detection: A Review of {DCASE}-Related Work",
    booktitle = dcase,
    year = "2025",
}

@inproceedings{guan2023time-weighted,
  author       = {Jian Guan and
                  Youde Liu and
                  Qiaoxi Zhu and
                  Tieran Zheng and
                  Jiqing Han and
                  Wenwu Wang},
  title        = {Time-Weighted Frequency Domain Audio Representation with {GMM} Estimator for Anomalous Sound Detection},
  booktitle    = icassp,
  year         = {2023}
}

@inproceedings{wu2025towards,
  title={Towards Few-Shot Training-Free Anomaly Sound Detection},
  author={Wu, Ho-Hsiang and Lin, Wei-Cheng and Kumar, Abinaya and Bondi, Luca and Ghaffarzadegan, Shabnam and Bello, Juan Pablo},
  booktitle=interspeech,
  year={2025}
}

@inproceedings{wilkinghoff2023using,
  author       = {Kevin Wilkinghoff and
                  Fabian Fritz},
  title        = {On Using Pre-Trained Embeddings for Detecting Anomalous Sounds with Limited Training Data},
  booktitle    = eusipco,
  year         = {2023},
}

@article{yadav2025overview,
  title={An overview of neural architectures for self-supervised audio representation learning from masked spectrograms},
  author={Yadav, Sarthak and Theodoridis, Sergios and Tan, Zheng-Hua},
  journal={arXiv:2509.18691},
  year={2025}
}

@article{radenovic2019fine-tuning,
  author       = {Filip Radenovic and
                  Giorgos Tolias and
                  Ondrej Chum},
  title        = {Fine-Tuning {CNN} Image Retrieval with No Human Annotation},
  journal      = ieee-tpami,
  volume       = {41},
  year         = {2019}
}

@article{nafchi2015deviation,
  author       = {Hossein Ziaei Nafchi and
                  Rachid Hedjam and
                  Atena Shahkolaei and
                  Mohamed Cheriet},
  title        = {Deviation Based Pooling Strategies For Full Reference Image Quality Assessment},
  journal      = {arXiv:1504.06786},
  year         = {2015},
}

@inproceedings{kolesnikov2016seed,
  author       = {Alexander Kolesnikov and
                  Christoph H. Lampert},
  title        = {Seed, Expand and Constrain: Three Principles for Weakly-Supervised Image Segmentation},
  booktitle    = eccv,
  year         = {2016},
}

@inproceedings{snyder2018x-vectors,
  author       = {David Snyder and
                  Daniel Garcia{-}Romero and
                  Gregory Sell and
                  Daniel Povey and
                  Sanjeev Khudanpur},
  title        = {X-Vectors: Robust {DNN} Embeddings for Speaker Recognition},
  booktitle    = icassp,
  year         = {2018}
}

@inproceedings{okabe2018attentive,
  author       = {Koji Okabe and
                  Takafumi Koshinaka and
                  Koichi Shinoda},
  title        = {Attentive Statistics Pooling for Deep Speaker Embedding},
  booktitle    = interspeech,
  year         = {2018}
}

@inproceedings{arandjelovic2018objects,
  author    = {Relja Arandjelovic and
               Andrew Zisserman},
  title     = {Objects that Sound},
  booktitle = eccv,
  year      = {2018}
}

@inproceedings{arandjelovic2017look,
  author    = {Relja Arandjelovic and
               Andrew Zisserman},
  title     = {Look, Listen and Learn},
  booktitle = iccv,
  year      = {2017}
}

@inproceedings{harada2021toyadmos2,
  author       = {Noboru Harada and
                  Daisuke Niizumi and
                  Daiki Takeuchi and
                  Yasunori Ohishi and
                  Masahiro Yasuda and
                  Shoichiro Saito},
  title        = {Toy{ADMOS2:} {A}nother Dataset of Miniature-Machine Operating Sounds for Anomalous Sound Detection under Domain Shift Conditions},
  booktitle    = dcase,
  year         = {2021}
}

@inproceedings{dohi2022mimiidg,
  author       = {Kota Dohi and others},
  title        = {{MIMII} {DG:} {S}ound Dataset for Malfunctioning Industrial Machine Investigation and Inspection for Domain Generalization Task},
  booktitle    = dcase,
  year         = {2022},
}

@inproceedings{purohit2019mimii,
  author       = {Harsh Purohit and others},
  title        = {{MIMII} Dataset: {S}ound Dataset for Malfunctioning Industrial Machine Investigation and Inspection},
  booktitle    = dcase,
  year         = {2019}
}

@inproceedings{koizumi2019toyadmos,
  author       = {Yuma Koizumi and
                  Shoichiro Saito and
                  Hisashi Uematsu and
                  Noboru Harada and
                  Keisuke Imoto},
  title        = {Toy{ADMOS}: {A} Dataset of Miniature-Machine Operating Sounds for Anomalous
                  Sound Detection},
  booktitle    = waspaa,
  year         = {2019}
}

@inproceedings{niizumi2024toyadmos2sharp,
    author = "Niizumi, Daisuke and Harada, Noboru and Ohishi, Yasunori and Takeuchi, Daiki and Yasuda, Masahiro",
    title = "{ToyADMOS2\#: Y}et Another Dataset for the {DCASE2024} Challenge Task 2 First-Shot Anomalous Sound Detection",
    booktitle = dcase,
    year = "2024"
}

@inproceedings{harada2025toyadmos2025,
    author = "Harada, Noboru and Niizumi, Daisuke and Ohishi, Yasunori and Takeuchi, Daiki and Yasuda, Masahiro",
    title = "{ToyADMOS2025}: The Evaluation Dataset for the {DCASE2025T2} First-Shot Unsupervised Anomalous Sound Detection for Machine Condition Monitoring",
    booktitle = dcase,
    year = "2025",
}

@article{saengthong2025retaining,
  author       = {Phurich Saengthong and
                  Tomoya Nishida and
                  Kota Dohi and
                  Natsuo Yamashita and
                  Yohei Kawaguchi},
  title        = {Retaining Mixture Representations for Domain Generalized Anomalous
                  Sound Detection},
  journal      = {arXiv:2510.25182},
  year         = {2025},
}

@article{fujimura2026pseudo,
  title={Pseudo-label distillation for discriminative
anomalous sound detection},
  author={Takuya Fujimura and Tomoki Toda},
  journal={arXiv preprint arXiv:2607.16678},
  year={2026},
}

@article{mcclish1989analyzing,
  author={McClish, Donna Katzman},
  title={{A}nalyzing a portion of the {ROC} curve},
  journal={Medical decision making},
  volume={9},
  number={3},
  year={1989},
}

@article{wang2022ranked,
  author       = {Xinshao Wang and
                  Yang Hua and
                  Elyor Kodirov and
                  Neil M. Robertson},
  title        = {Ranked List Loss for Deep Metric Learning},
  journal      = ieee-tpami,
  volume       = {44},
  number       = {9},
  year         = {2022}
}

@inproceedings{mezza2023zero-shot,
  author       = {Alessandro Ilic Mezza and
                  Giulio Zanetti and
                  Maximo Cobos and
                  Fabio Antonacci},
  title        = {Zero-Shot Anomalous Sound Detection in Domestic Environments Using Large-Scale Pretrained Audio Pattern Recognition Models},
  booktitle    = icassp,
  year         = {2023}
}

@inproceedings{wilkinghoff2024adaproj,
   author = {Wilkinghoff, Kevin},
   title = {Ada{P}roj: {A}daptively Scaled Angular Margin Subspace Projections for Anomalous Sound Detection with Auxiliary Classification Tasks},
   booktitle = dcase,
   year = {2024}
}

@inproceedings{harada2023toyadmos2+,
   author = "Harada, Noboru and Niizumi, Daisuke and Takeuchi, Daiki and Ohishi, Yasunori and Yasuda, Masahiro",
   title = "{ToyADMOS2+}: New {T}oyadmos Data and Benchmark Results of the First-Shot Anomalous Sound Event Detection Baseline",
   booktitle = dcase,
   year = "2023"
}

@inproceedings{albertini2024imadds,
   author = "Albertini, Davide and Augusti, Filippo and Esmer, Kudret and Bernardini, Alberto and Sannino, Roberto",
   title = "{IMAD-DS}: A Dataset for Industrial Multi-Sensor Anomaly Detection Under Domain Shift Conditions",
   booktitle = dcase,
   year = "2024"
}

@inproceedings{jiang2025adaptive,
  title={Adaptive Prototype Learning for Anomalous Sound Detection with Partially Known Attributes},
  author={Jiang, Anbai and others},
  booktitle=icassp,
  year={2025},
}

@inproceedings{dinkel2024dasheng,
  author={Dinkel, Heinrich and Yan, Zhiyong and Wang, Yongqing and Zhang, Junbo and Wang, Yujun and Wang, Bin},
  title={Scaling up masked audio encoder learning for general audio classification},
  booktitle=interspeech,
  year={2024}
}

@inproceedings{wilkinghoff2023design,
  author = {Wilkinghoff, Kevin},
  title = {Design Choices for Learning Embeddings from Auxiliary Tasks for Domain Generalization in Anomalous Sound Detection},
  booktitle = icassp,
  year = {2023},
}

@inproceedings{liu2022anomalous,
  author       = {Youde Liu and
                  Jian Guan and
                  Qiaoxi Zhu and
                  Wenwu Wang},
  title        = {Anomalous Sound Detection Using Spectral-Temporal Information Fusion},
  booktitle    = icassp,
  year         = {2022}
}

@inproceedings{harada2023first,
  author       = {Noboru Harada and
                  Daisuke Niizumi and
                  Yasunori Ohishi and
                  Daiki Takeuchi and
                  Masahiro Yasuda},
  title        = {{F}irst-Shot Anomaly Sound Detection for Machine Condition Monitoring: {A} Domain Generalization Baseline},
  booktitle    = eusipco,
  year         = {2023}
}

@inproceedings{hou2023decoupling,
  author       = {Qijun Hou and
                  Anbai Jiang and
                  Wei{-}Qiang Zhang and
                  Pingyi Fan and
                  Jia Liu},
  title        = {Decoupling Detectors for Scalable Anomaly Detection in {AIoT} Systems with Multiple Machines},
  booktitle    = {Proc. GLOBECOM},
  year         = {2023}
}

@inproceedings{wilkinghoff2024self,
  author = {Wilkinghoff, Kevin},
  title = {Self-Supervised Learning for Anomalous Sound Detection},
  booktitle   = icassp,
  year   = {2024},
}

@inproceedings{dawalatabad2021ecapa-tdnn,
  author       = {Nauman Dawalatabad and
                  Mirco Ravanelli and
                  Fran{\c{c}}ois Grondin and
                  Jenthe Thienpondt and
                  Brecht Desplanques and
                  Hwidong Na},
  title        = {{ECAPA-TDNN} Embeddings for Speaker Diarization},
  booktitle    = interspeech,
  year         = {2021}
}

@inproceedings{han2024exploring,
  author       = {Bing Han and others},
  title        = {Exploring Large Scale Pre-Trained Models for Robust Machine Anomalous Sound Detection},
  booktitle    = icassp,
  year         = {2024}
}

@inproceedings{zhang2024dual,
  author       = {Yucong Zhang and
                  Juan Liu and
                  Yao Tian and
                  Haifeng Liu and
                  Ming Li},
  title        = {A Dual-Path Framework with Frequency-and-Time Excited Network for
                  Anomalous Sound Detection},
  booktitle    = icassp,
  year         = {2024}
}

@inproceedings{yin2025diffusion,
  title={Diffusion Augmentation Sub-center Modeling for Unsupervised Anomalous Sound Detection with Partially Attribute-Unavailable Conditions},
  author={Yin, Jiawei and Gao, Yu and Zhang, Wenbin and Wang, Tianyi and Zhang, Mingjun},
  booktitle=icassp,
  year={2025},
}

@inproceedings{yin2024self,
  title={Self-Supervised Augmented Diffusion Model for Anomalous Sound Detection},
  author={Yin, Jiawei and Zhang, Wenbin and Zhang, Mingjun and Gao, Yu},
  booktitle=apsipa,
  year={2024},
}

@article{han2025exploring,
  title={Exploring Self-Supervised Audio Models for Generalized Anomalous Sound Detection},
  author={Han, Bing and Jiang, Anbai and Zheng, Xinhu and Zhang, Wei-Qiang and Liu, Jia and Fan, Pingyi and Qian, Yanmin},
  journal=ieee-taslp,
  year={2025},
  volume={33}
}

@inproceedings{zhang2019mixup,
  author       = {Hongyi Zhang and
                  Moustapha Ciss{\'{e}} and
                  Yann N. Dauphin and
                  David Lopez{-}Paz},
  title        = {Mixup: Beyond Empirical Risk Minimization},
  booktitle    = iclr,
  year         = {2018},
}

@inproceedings{zhang2025echo,
  title={{ECHO}: Frequency-aware Hierarchical Encoding for Variable-length Signals},
  author={Yucong Zhang and Juan Liu and Ming Li},
  booktitle=icassp,
  year={2026},
}

@article{fan2025fisher,
  title={{FISHER}: A Foundation Model for Multi-Modal Industrial Signal Comprehensive Representation},
  author={Fan, Pingyi and others},
  journal={arXiv preprint arXiv:2507.16696},
  year={2025}
}

@article{kong2020sound,
  author       = {Qiuqiang Kong and
                  Yong Xu and
                  Wenwu Wang and
                  Mark D. Plumbley},
  title        = {Sound Event Detection of Weakly Labelled Data With {CNN}-Transformer and Automatic Threshold Optimization},
  journal      = ieee-acm-taslp,
  volume       = {28},
  pages        = {2450--2460},
  year         = {2020}
}

@techreport{giri2020unsupervised,
    Author = "Giri, Ritwik and Tenneti, Srikanth V. and Helwani, Karim and Cheng, Fangzhou and Isik, Umut and Krishnaswamy, Arvindh",
    title = "Unsupervised Anomalous Sound Detection Using Self-Supervised Classification and Group Masked Autoencoder for Density Estimation",
    institution = "DCASE2020 Challenge",
    year = "2020"
}

@techreport{zeng2022robust,
    Author = "Zeng, Ying and Liu, Hongqing and Xu, Lihua and Zhou, Yi and Gan, Lu",
    title = "Robust Abomaly Sound Detection Framework for Machine Condition Monitoring",
    institution = "DCASE2022 Challenge",
    year = "2022"
}

@techreport{jie2023anomalous,
    Author = "Junjie, Wang and Jiajun, Wang and Shengbing, Chen and Yong, Sun and Mengyuan, Liu",
    title = "Anomalous Sound Detection Based on Self-Supervised Learning",
    institution = "DCASE2023 Challenge",
    year = "2023"
}

@techreport{lv2024aithu,
    Author = "Lv, Zhiqiang and others",
    title = "{AITHU} System for First-Shot Unsupervised Anomalous Sound Detection",
    institution = "DCASE2024 Challenge",
    year = "2024",
}

@techreport{wang2025pre-trained,
    Author = "Wang, Lei",
    title = "Pre-Trained Model Enhanced Anomalous Sound Detection System for {DCASE2025} Task2",
    institution = "DCASE2025 Challenge",
    year = "2025"
}

@inproceedings{yadav2024audio,
  author       = {Sarthak Yadav and
                  Zheng{-}Hua Tan},
  title        = {Audio Mamba: Selective State Spaces for Self-Supervised Audio Representations},
  booktitle    = interspeech,
  year         = {2024}
}

@inproceedings{baevski2022data2vec,
  author       = {Alexei Baevski and
                  Wei{-}Ning Hsu and
                  Qiantong Xu and
                  Arun Babu and
                  Jiatao Gu and
                  Michael Auli},
  title        = {data2vec: {A} General Framework for Self-supervised Learning in Speech, Vision and Language},
  booktitle    = icml,
  year         = {2022},
}

@article{chen2022wavlm,
  author       = {Sanyuan Chen and others},
  title        = {{WavLM}: Large-Scale Self-Supervised Pre-Training for Full Stack Speech Processing},
  journal      = {{IEEE} J. Sel. Top. Signal Process.},
  year         = {2022},
}

@article{liu2022audio,
  author       = {Shuo Liu and others},
  title        = {Audio self-supervised learning: {A} survey},
  journal      = {Patterns},
  volume       = {3},
  number       = {12},
  year         = {2022}
}

@inproceedings{gemmeke2017audioset,
  author       = {Jort F. Gemmeke and others},
  title        = {Audio Set: An ontology and human-labeled dataset for audio events},
  booktitle    = icassp,
  year         = {2017},
}

\vfill

\end{document}